# A Comparison of the Red and Green Coronal Line Intensities at the 29 March 2006 and the 1 August 2008 Total Solar Eclipses: Considerations of the Temperature of the Solar Corona

A. Voulgaris

*Evrivias 5c, GR-542 50 Thessaloniki, Greece*

T. Athanasiadis, J.H. Seiradakis

*Aristotle University of Thessaloniki, Department of Physics, Section of Astrophysics, Astronomy and Mechanics, GR-541 24 Thessaloniki, Greece*

J.M. Pasachoff

*Williams College-Hopkins Observatory, Williamstown, Massachusetts 01267, USA*

During the total solar eclipse at Akademgorodok, Siberia, Russia, on 1 August 2008, we imaged the flash spectrum with a slitless spectrograph. We have spectroscopically determined the duration of totality, the epoch of the 2$^{nd}$ and 3$^{rd}$ contacts and the duration of the flash spectrum. Here we compare the 2008 flash spectra with those that we similarly obtained from the total solar eclipse of 29 March 2006, at Kastellorizo, Greece. Any changes of the intensity of the coronal emission lines, in particularly those of [Fe X] and [Fe XIV], could give us valuable information about the temperature of the corona. The results show that the ionization state of the corona, as manifested especially by the [Fe XIV] emission line, was much weaker during the 2008 eclipse, indicating that following the long, inactive period during the solar minimum, there was a drop in the overall temperature of the solar corona.

## 1. Introduction

We have studied, with a slitless spectrograph, the flash spectrum emission lines of [Fe X] and [Fe XIV], during two solar eclipses: the total solar eclipse of 29 March 2006 observed at the island of Kastellorizo, Greece, and the total solar eclipse of 1 August 2008 at Akademgorodok (near Novosibirsk), Siberia, Russia (Pasachoff et al., 2007, 2009). The Sun was fairly active during the 2006 eclipse. In contrast, it was quiet and inactive during the 2008 eclipse, and for a very long interval before the eclipse. Any changes of the emission lines between the two epochs could be directly attributed to changes in the activity of the Sun (Esser et al., 1995; Kanno,

Tsubaki and Kurokawa, 1971; Singh, Gupta and Cowsik, 1997; Brickhouse, Esser and Habbal, 1995; Singh et al., 1999; Pasachoff, 2009a; Pasachoff, 2009b, Golub and Pasachoff, 2009).

## 2. The 1 August 2008 total solar eclipse at Akademgorodok, Russia

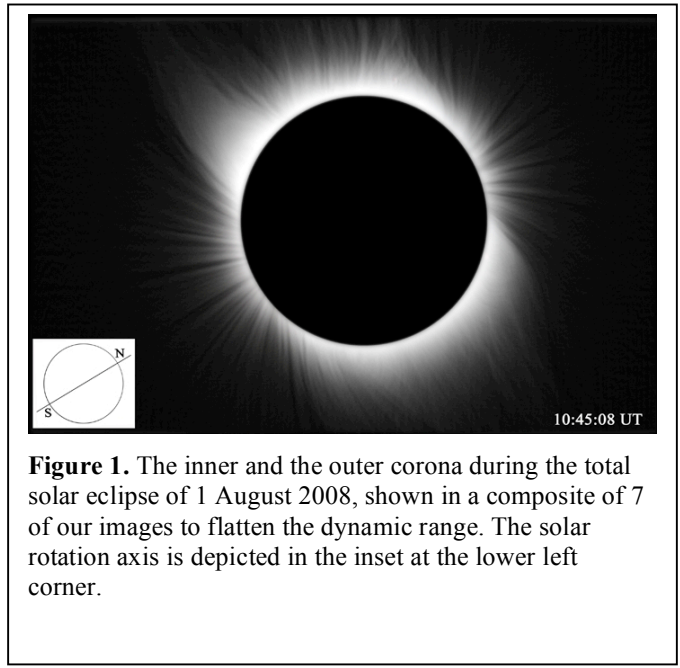

**Figure 1.** The inner and the outer corona during the total solar eclipse of 1 August 2008, shown in a composite of 7 of our images to flatten the dynamic range. The solar rotation axis is depicted in the inset at the lower left corner.

The observations of the 2008 total solar eclipse described herein (Figure 1) were undertaken at the roof of the Budker Institute of Nuclear Physics, in Akademgorodok, near Novosibirsk, in southern Siberia, Russia. At that location ($\lambda = 83º\ 06'\ 43''.8$ E, $\varphi = +54º\ 50'\ 57''.1$), the maximum of the eclipse occurred at UT $10^h 45^m 36^s.5$.

2.1. The slitless spectrograph used during the 2008 eclipse

In order to record the flash spectrum, a slitless spectrograph with a reflection grating of 300 lines/mm, blazed at 5000 Å, and an 135 mm f/3.5 telephoto lens was constructed by one of us (A.V.).

The system's efficiency was about 60% at 5303 Å (Fe XIV emission line) and 20% at 6374 Å (Fe X emission line). The grating was placed before the telephoto lens on a rotating turntable that allowed the rotation of the grating. The full range of the visible spectrum, from 3900 Å to 6700 Å, was projected on the CCD sensor of the digital camera. The resolution of the spectrograph was 1.5 Å /pixel. The diameter of the Sun corresponded to 275 pixels or 6.87 ″/pixel. With the help of the turntable, the direction of the grating lines could be set parallel to the direction of the last visible elongated crescent of the Sun; this narrow crescent played the role of the "slit" in the slitless spectrograph. The solar light was directed to the spectrograph by a simple reflection coelostat (Veio, 1977; Dougherty, 1982; Demianski and Pasachoff 1984; Pasachoff and Livingston, 1984), which was also constructed by one of us (A.V.).

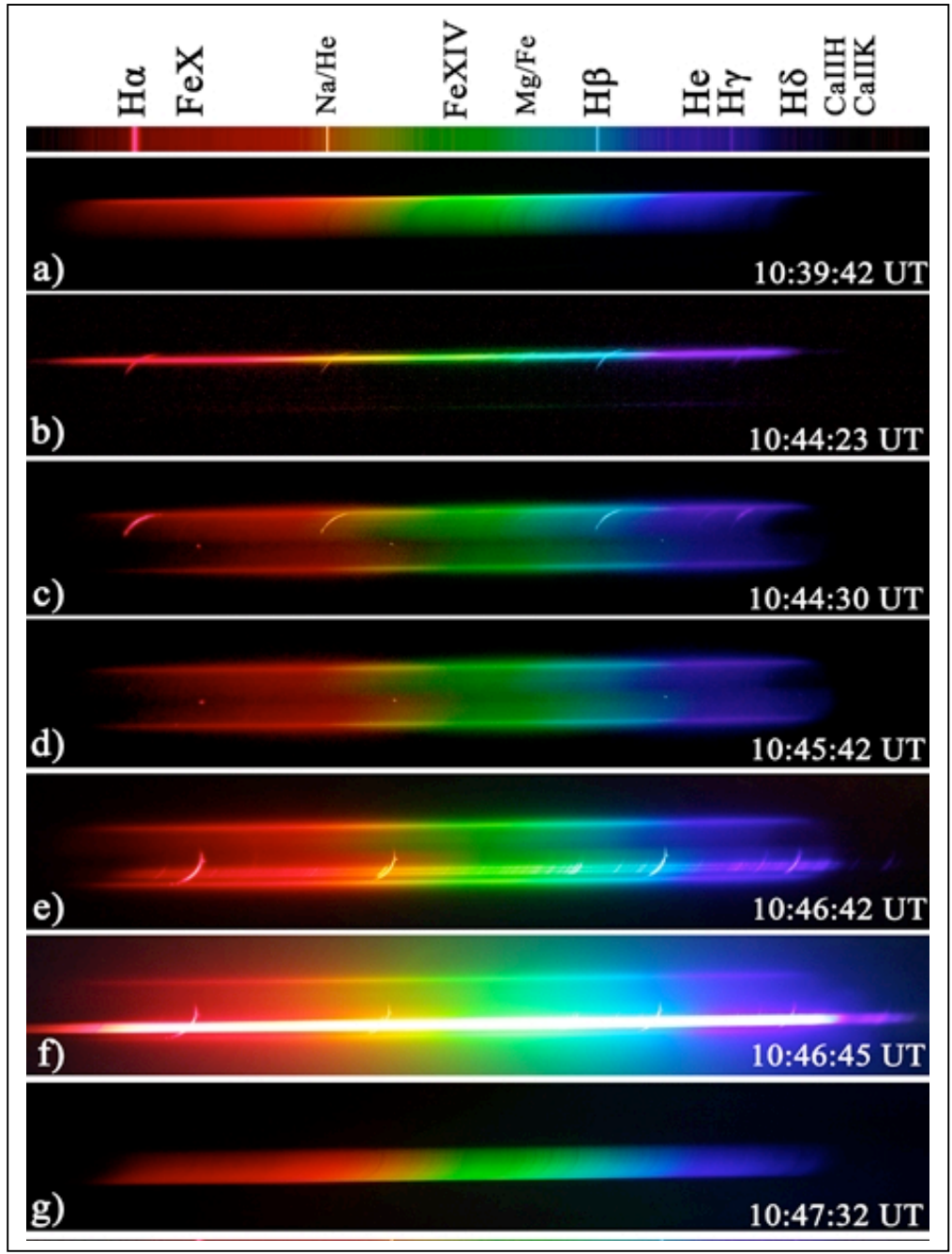

## 2.2. The flash spectrum of the total solar eclipse of 1 August 2008

Data were recorded during the ingress and egress phases. However, the ingress data were underexposed. They were mainly used for adjusting the exposure time of the egress phase. They proved very useful for estimating the duration of totality and the exact time of 2$^{nd}$ and 3$^{rd}$ contact. A sequence of 91 spectra (some of which are presented in Figure 2) were recorded during totality and the partial phases, with ~1/50 s exposure time.

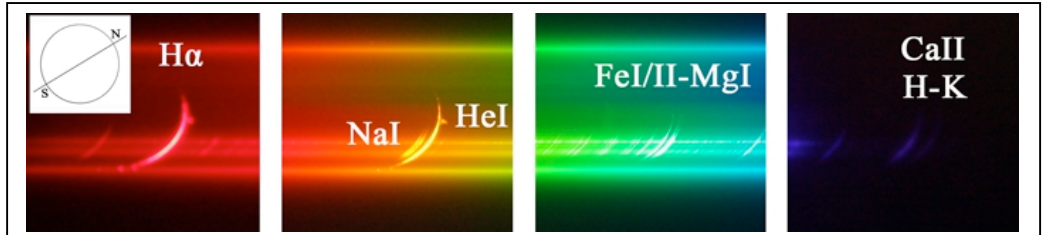

**Figure 3.** Snapshots of flash-spectrum emission lines from the chromosphere and a prominence.

Among the chromospheric emission lines observed (Figure 3) in the flash spectrum, were emission lines from hydrogen (6562.8 Å), sodium (5890 Å, 5896 Å), helium 5876 Å (the $D_3$ discovery line), magnesium I "b" (5167.3 Å, 5172.7 Å, 5183.6 Å), neutral and once-ionized iron (5168.9 Å, 5169.1 Å – blended with the magnesium lines) and the H (3968.5 Å) and K (3933.6 Å) lines of once-ionized calcium. (For the origin of the K-line notation, see Pasachoff and Suer, 2009, 2010.)

Of the two characteristic coronal emission lines that are produced by the multiply ionized iron in the visible part of the spectrum (Figure 5), the [Fe XIV] (5302.8 Å) coronal emission line, which requires a temperature of about $1.9 \times 10^6$ K, was barely visible, whereas the much lower-temperature ($1.2 \times 10^6$ K) [Fe X] (6374 Å) line was clearly visible (Figure 4) (Singh et al., 2002; Takeda et al., 2000).

2.3. Spectroscopic definition of $2^{nd}$ and $3^{rd}$ contact, totality and the duration of the 1 August 2008 flash spectrum

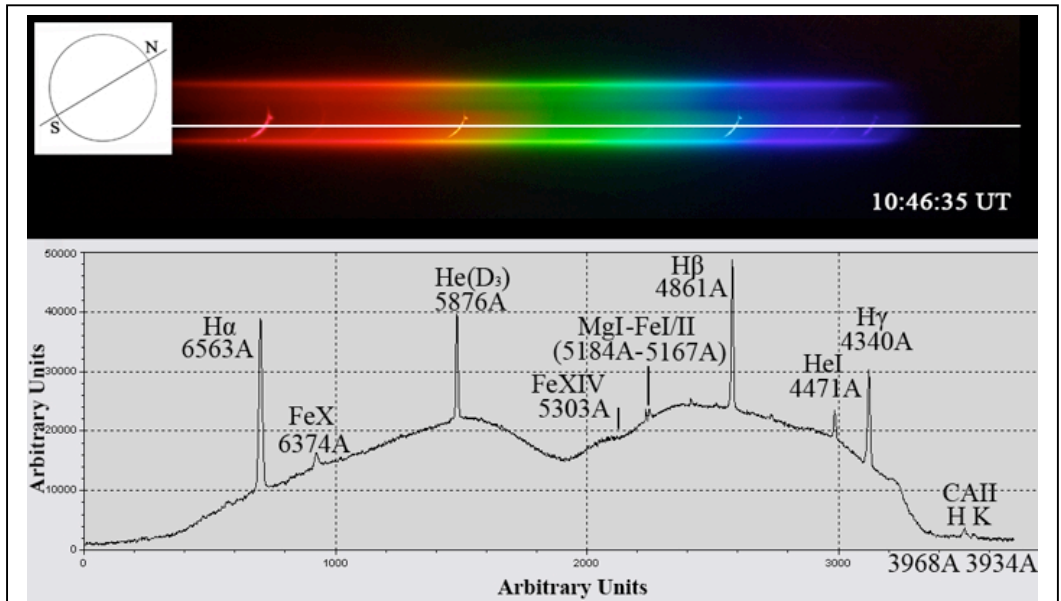

**Figure 4.** The flash spectrum of the 1 August 2008 total solar eclipse (top) and a cut through it, plotted on a linear scale (bottom). The cut in the bottom panel was taken along the white line in the top panel. The moderate intensity of the [Fe X] emission line is clearly visible. The extremely weak emission of the [Fe XIV] line is hardly visible.

2.3.1 Spectroscopic definition of the duration of the flash spectrum, the 2$^{nd}$ and 3$^{rd}$ contacts, and the duration of totality, using the Hα emission line.

The duration of the flash spectrum depends on the geometrical and physical characteristics of the investigated eclipse. In particular, it depends on the angular velocity of the Moon (relatively to the Sun) and the height of the chromosphere, which depends on solar activity. Statistics of spicule height and other parameters were recently re-measured by Pasachoff, Jacobson, and Sterling (2009). Furthermore, the *ingress* and *egress* events are symmetrical only when the center of the disks of the Sun and the Moon coincide during the middle of the eclipse.

Until 10:43:55 UT, the solar spectrum was dominated by absorption lines (Figure 2a). For the next 25 s, both absorption and emission lines were present (clearly seen in Figure 5). At 10:44:20 UT the absorption lines disappeared. At 10:44:30.5±0.5 UT, the maximum Hα emission was observed, corresponding to the spectroscopic Hα 2$^{nd}$ contact (Figure 2c). During the next 28 seconds, the intensity of the Hα emission line decreased (while the chromosphere on the east limb was being eclipsed). Between 10:44:58 UT and 10:46:21 UT, no chromospheric Hα emission was observed (but there was strong Hα and He emission from intense prominences in both the east and west limbs). The middle of the eclipse (assumed to be symmetrically centered between 2$^{nd}$ and 3$^{rd}$ contacts) occurred at 10:45:36±2 UT. After totality, chromospheric Hα emission increased (when the chromosphere on the west limb was emerging from eclipse), reaching a maximum at 10:46:43±1 UT (Figure 2e), indicating the spectroscopic Hα 3$^{rd}$ contact. For the next 25 s, Hα emission (Figure 2f) gradually decreased, and at 10:47:12 UT, the solar spectrum was again dominated by absorption lines (Figure 2g).

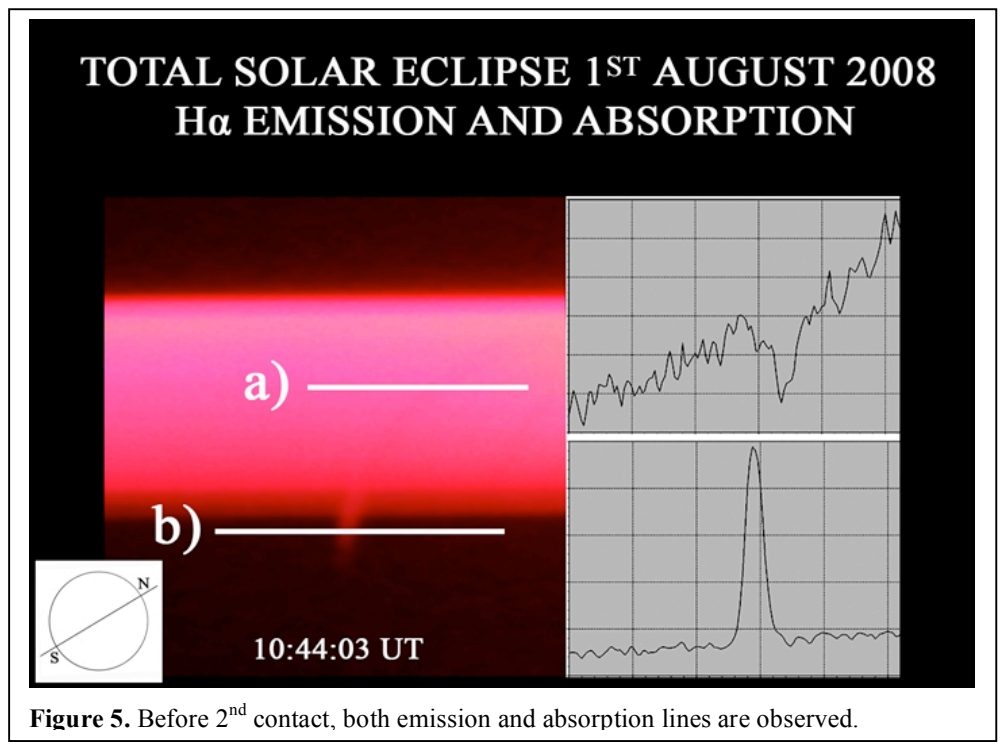

**Figure 5.** Before 2$^{nd}$ contact, both emission and absorption lines are observed.

In Figure 6, the intensity of Hα, Hα = $I_\lambda - I_c$ (where, $I_\lambda$ is the integrated intensity underneath the Hα line profiles and $I_c$ is the corresponding continuum intensity interpolated either side of the line emission) is presented as a function of time. The intensities were calculated by integrating the data underneath the line profiles. The intensity of the continuum emission was calculated by interpolating the continuum intensity either side of the line emission.

During ingress (east limb of the Sun), Hα and He emission lines were observed in our spectra for 63±2 seconds. During egress (west limb), this interval became smaller, 48±3 seconds (Figure 8). It is not surprising that the two durations are not equal, as the eclipse was not centrally symmetric. The duration of the eclipse (calculated between the two Hα peaks) according to the above measurements was $2^m$ $12^s$. During the 83 middle seconds of this interval, no Hα emission was detected (Figure 6).

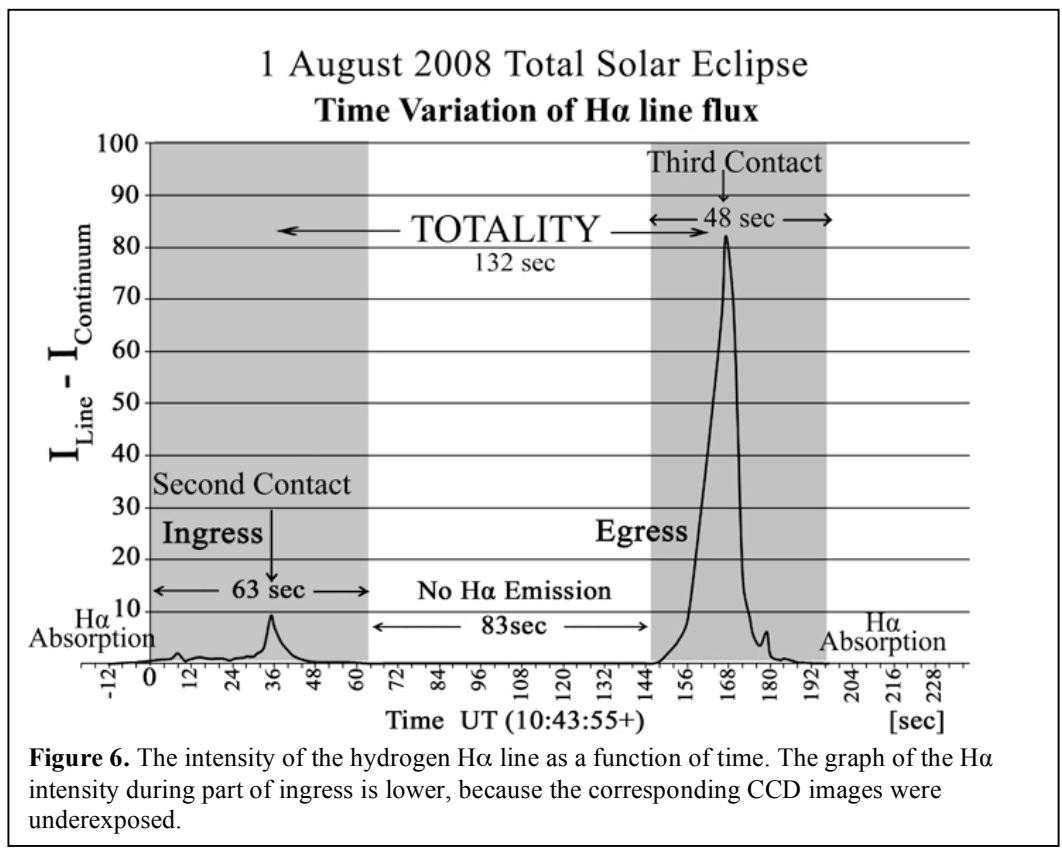

**Figure 6.** The intensity of the hydrogen Hα line as a function of time. The graph of the Hα intensity during part of ingress is lower, because the corresponding CCD images were underexposed.

The online manifestation of Espenak's calculations (http://eclipse.gsfc.nasa.gov/SEmono/TSE2008/TSE2008tab/TSE2008-Table13.pdf) by Xavier Jubier (http://xjubier.free.fr/en/site_pages/SolarEclipseCalc_Diagram.html), shows that at Akademgorodok the $2^{nd}$ contact occurred at 10:44:26.0 UT and the $3^{rd}$ contact at 10:46:44.3 UT, with mid-eclipse at 10:45:36.5 UT and a total duration of 2 min 17.3 s. Thus, the spectroscopic Hα measurements are 4.5±0.5 s later than the prediction of Jubier for the $2^{nd}$ contact and 1.3±1 s earlier than the prediction for the $3^{rd}$ contact. The duration of the eclipse is 5.3 s shorter than Espenak's calculations, which is significant. However the mid eclipse differs by only 0.5±2 s, which, in effect, coincides with Espenak's calculations.

2.3.2 Spectroscopic definition of the 2$^{nd}$ and 3$^{rd}$ contacts, and the duration of totality, using the Mg I and Fe I/II spectral emission lines.

Among the most characteristic heavy-element emission lines, observed in the solar chromosphere, are those of Mg I b lines (5167.3Å, 5172.7 Å, 5183.6 Å), Fe I (5168.9 Å) and Fe II (5169.1 Å) in the green part of the spectrum. These are generated by elements in the lower chromosphere. In contrast, the hydrogen Hα and helium $D_3$ lines are emitted over a larger height range, extending well above the photosphere. The thin layer emitting the Mg I and Fe I/II lines can be used to identify the limb of the solar disk. Therefore the 2$^{nd}$ contact can be considered to correspond to the disappearance of the magnesium and iron emission lines and the 3$^{rd}$ contact to the reappearance of these lines. We observed the disappearance and the reappearance of these lines during the 1 August 2008 eclipse. According to the above consideration, the spectroscopically defined 2$^{nd}$ contact occurred at 10:44:32 ± 1 s UT and the 3$^{rd}$ contact at 10:46:33 ± 2 s UT, giving a duration of 2 m 01 s ± 2 s, centered at 10:45:32.5 ± 2 s UT. We note that the duration of the totality, defined by the Mg I and Fe I/II lines was shorter by 17.3 ±2 s than the photospheric duration calculated by Espenak (Espenak and Anderson, 2007).

## 3. The 29 March 2006 total solar eclipse at Kastellorizo, Greece

The observations of the total solar eclipse were undertaken at the patio of the Hotel Megisti (λ = 29º 35′ 28″.8, φ = +36º 09′ 08″.6) on the island of Kastellorizo. At that location the maximum of the eclipse occurred at $10^h53^m28^s.1$ UT.

3.1. The slitless spectrograph used during the 2006 eclipse

A holographic transmission grating, with 566 lines/mm, used with an f/3.5 135 mm telephoto lens, was used for the recording of the flash spectrum during the total solar eclipse of 29 March 2006 at Kastellorizo. The system's efficiency was about 3% at 5303 Å (Fe XIV emission line) and 2% at 6374 Å (Fe X emission line). The lower efficiency of this system does not affect the results but the exposure time had to be increased to ~1 s. We also used a digital Canon EOS 350D camera. The full range of the visible spectrum, from 3950 Å to 6700 Å, was projected on the CCD sensor of the digital camera. The resolution of the spectrograph was 0.9 Å /pixel. The diameter of the Sun corresponded to 285 pixels or 6.53″/pixel. Spectra were recorded only during ingress. No data were recorded during egress.

3.2. The flash spectrum of the total solar eclipse of 29 March 2006

The flash spectrum of the total solar eclipse of 29 March 2006 was recorded a few seconds before totality (Figure 7). Four CCD images were recorded, showing bright emission lines corresponding to chromospheric H and He. The [Fe XIV] green line, at 5303 Å was clearly visible, whereas the emission of the [Fe X] red line was almost absent.

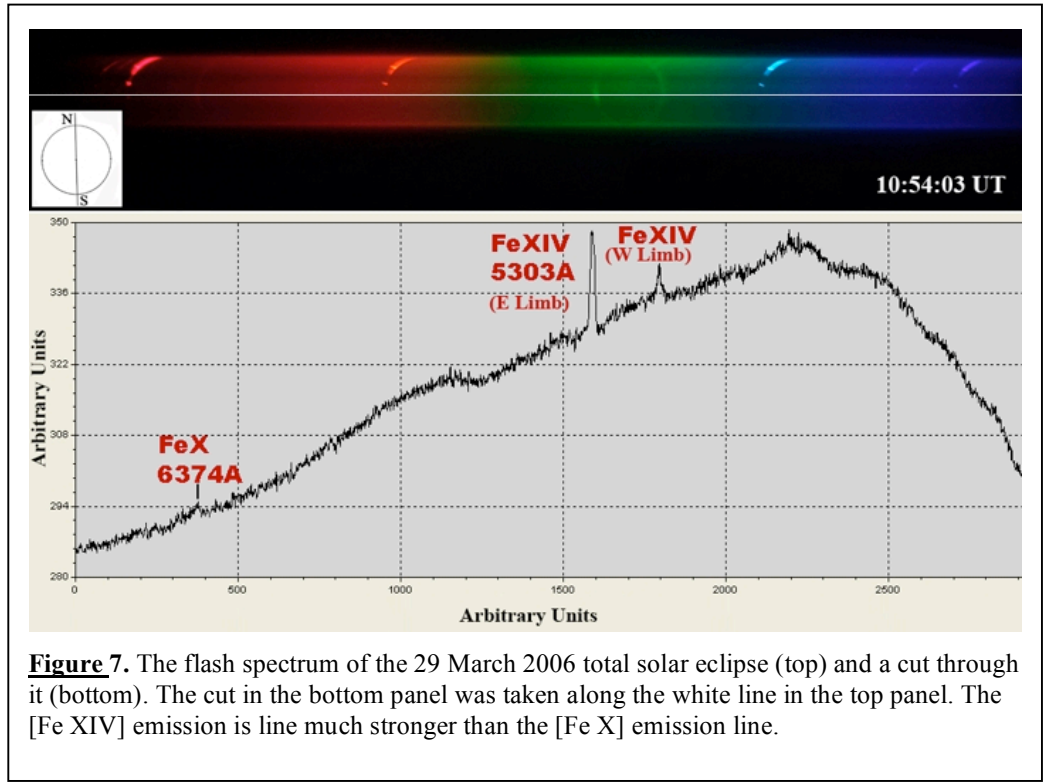

**Figure 7.** The flash spectrum of the 29 March 2006 total solar eclipse (top) and a cut through it (bottom). The cut in the bottom panel was taken along the white line in the top panel. The [Fe XIV] emission is line much stronger than the [Fe X] emission line.

### 4. Comparison of the 2006 and 2008 flash spectra

The [Fe X] and [Fe XIV] lines show obvious changes between the 2006 and 2008 eclipses. In particular the [Fe X] red line was much stronger in the 2008 spectra, whereas the [Fe XIV] green line was much stronger in the 2006 spectra (Figure 8). On the other hand, the chromospheric Hα and Hβ lines on the same spectra in Figure 8 are reasonably similar, which indicates that the above mentioned differences in the Fe emission lines were genuine.

### 5. Measurements and analysis

Initially, the original raw data were converted to FITS data, using the *IRIS* program (Buil, 2008). In order to be consistent, the calculations that follow were made at the same solar latitude for both the 2006 and the 2008 data. The orientation of the rotation axis of the Sun (see Figure 8, white lines) was estimated using the Observatoire de Paris – Meudon: 2010, *Bass 2000 Solar Survey Archive*, grid lines and spectroheliographic images. The alignment was achieved by overlaying the *Bass 2000* (2010) grid images on our spectra, using the prominences that were present at the time of the eclipse. However, either because the diffractive efficiency of the ruled gratings decreases with increasing deviation from their principal direction or because the intensity of the [Fe XIV] emission declines with distance from the solar equator, the useful range was limited between heliographic latitude 8°N and 28°N (±0.3°). No active regions were present in this range or its vicinity, in both 2006 and 2008 eclipses. Within this range, the integrated intensity over the emission line profile of [Fe X] and [Fe XIV] in the images was calculated as well as above the continuum background. This was done for the east limb for the 2006 eclipse and for the west limb for the 2008 eclipse.

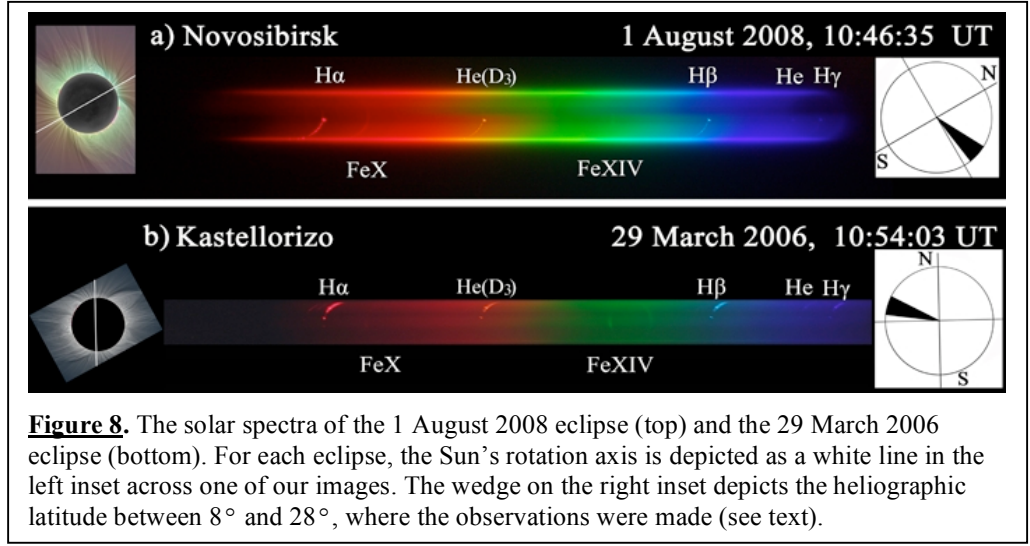

**Figure 8.** The solar spectra of the 1 August 2008 eclipse (top) and the 29 March 2006 eclipse (bottom). For each eclipse, the Sun's rotation axis is depicted as a white line in the left inset across one of our images. The wedge on the right inset depicts the heliographic latitude between 8° and 28°, where the observations were made (see text).

After the alignment of the spectra, the relative intensity, $I_{Fe}$, was calculated according to the formula
$$I_{Fe} = (I_\lambda - I_c)/I_c$$
where $I_\lambda$ is the integrated intensity of each forbidden Fe coronal line and $I_c$ is the corresponding intensity of the continuum background.

For the 2006 eclipse, the relative intensity for the [Fe XIV] line was $I_{Fe\ XIV}$ = 0.039±0.001 whereas for the [Fe X] line was $I_{FeX}$ = 0.022±0.001. Therefore, the ratio of the relative intensities was
$$I_{Fe\ X}/I_{Fe\ XIV} = 0.58 \pm 0.01$$

This ratio shows that the emission line of [Fe XIV] was much brighter than the corresponding [Fe X] line during the 2006 total solar eclipse at Kastellorizo.

For the 2008 eclipse, the relative intensity for the [Fe XIV] line was $I_{Fe\ XIV}$ = 0.024±0.001, whereas for the [Fe X] line was $I_{FeX}$ = 0.053±0.001. Therefore, the ratio of the relative intensities was
$$I_{Fe\ X}/I_{Fe\ XIV} = 2.16 \pm 0.05$$
This ratio shows that the emission line of [Fe XIV] was much weaker than the corresponding [Fe X] line during the 2008 total solar eclipse at Novosibirsk (Table 1).

**Table 1.** The relative intensities of the ionized iron lines for the 2006 and 2008 eclipses

| Emission lines | 2006, Kastellorizo Relative intensity | 2008, Novosibirsk Relative intensity |
|---|---|---|
| [Fe X] | 0.022 | 0.053 |
| [Fe XIV] | 0.039 | 0.024 |
| [Fe X]/ [Fe XIV] | 0.58 | 2.16 |

Finally, we compare the ratio of the intensities of the [Fe X] and [Fe XIV] emission lines for the above eclipses. The results are presented in Table 2, where it becomes obvious that the relative intensity of [Fe X] to [Fe XIV] in the solar corona has increased substantially between 2006 and 2008.

Table 2: Ratio of the relative intensities of the [Fe X] and [Fe XIV] lines between 2006 and 2008

| | |
|---|---|
| [Fe X](Novosibirsk)/[Fe X](Kastellorizo) | 2.38 |
| [Fe XIV](Novosibirsk)/[Fe XIV](Kastellorizo) | 0.63 |

## 6. Conclusions

As shown in Table 2, during the 2008 eclipse observed from Novosibirsk, the [Fe XIV] emission was much weaker than the [Fe X] emission. In contrast, during the 2006 eclipse observed from Kastellorizo, it was stronger. It should be noted that in the 1999 eclipse data, presented by Buil (2008), the [Fe XIV] line was also much stronger. Such changes could be attributed to the specific locations, where the data were taken in each observation. However, the continuous decline of the [Fe XIV] line intensity between 1999 and 2008 could be attributed to the decline of solar activity between 1999 and 2008. Solar activity was at a prolonged minimum during the 2008 eclipse, with sunspot number 0, whereas the sunspot number was about 20 during the period of the 2006 eclipse (still not high but certainly higher than 0) and 75 during the period of the 1999 eclipse. According to Esser et al. (1995), the intensities of the multiply ionized iron lines depend on the temperature of the corona (see their Figure 1b), so we are tempted to suppose that the decline of the [Fe XIV] emission was due to the decline of the corona temperature during the last two years (2006–2008). This, in turn, could be attributed to the very long and overdue minimum of the solar activity, which was reflected by the 345 (out of 856) spotless days between the 2006 and the 2008 total solar eclipses.

According to Nikolskaia and Utrobin (1984) the low intensity of the [Fe XIV] emission indicates that the temperature of the observed area must be less than $1.9 \times 10^6$ K. Similarly, the intensity of the [Fe X] emission line requires a corona temperature larger than $1.2 \times 10^6$ K. It is noted that the decrease of the [Fe XIV] emission line, could be attributed to either an increase or a decrease of the solar temperature. However, the absence of the high temperature lines of [Ni XV] 6702 Å ($2.3 \times 10^6$ K), [Ca XIII] 4086 Å ($2.3 \times 10^6$ K) and [Ca XV] 5694 Å ($3.8 \times 10^6$ K) justifies our assumption that the decrease is due to a lower corona temperature (and not a higher one) (Halas et al., 1997), as is already obvious from the intensity of X-ray emission.

Esser et al. (1995 – see their Figure 1) used two independent models (by Brickhouse et al., 1995 and by Mason, 1975), to calculate the temperature of the solar corona as a function of the ratio of the [Fe XIV]/[Fe X] line intensities. Using the [Fe XIV]/[Fe X] data presented in Table 1, and the graphs of the above mentioned figure, the temperature of the regions of the solar corona that we analyzed can be estimated: According to the Brickhouse et al., (1995) model, the drop of the temperature between 2006 and 2008 was $(1.1\pm0.1)\times10^5$ K. According to the Mason (1975) model it was a little larger, $(1.3\pm0.1)\times10^5$ K. Mazzotta et al. (1998) present newer calculations of the ionization equilibrium for [Fe X] and for [Fe XIV]. Comparing our [Fe X]/[Fe XIV] estimates with the data presented in their Table 2, the drop of the temperature of the solar corona between 2006 and 2008 was $(3.0\pm0.5)\times10^5$ K. Similar calculations of the relation of the ionic stage to temperature were carried out by J. Raymond (quoted in Golub and Pasachoff, 2009).

Habbal et al. (2010) discuss the inference of electron temperature from emission lines, and the change of emission from collisional to radiative domination. Our observations were in the inner, collisional region. Habbal et al. (2010) and also Habbal et al. (2009) and Daw et al. (2010) provide 2006 and 2008 eclipse observations of the [Fe XI] infrared line at 7892 Å. The latter paper provides a two-dimensional spatial comparison of the [Fe XI] specified as being similar to [Fe X], [Fe XIII], [Fe XIV] regions of prime emission, extending much higher in the corona than the regions shown in our spectra.

As discussed above, we attribute the drop of the temperature to the observed prolonged minimum of solar activity.

The duration of the flash spectrum was spectroscopically estimated from the appearance of the Hα emission line, which was observed during the ingress and during the egress phases of the 2008 eclipse.

Finally the disappearance and later reappearance of the neutral magnesium "b lines" (5167.3Å, 5172.7 Å, 5183.6 Å) and the neutral (5168.9 Å) and the once-ionized iron (5169.1 Å) emission lines of the thin lower chromosphere were used to spectroscopically define the 2$^{nd}$ and 3$^{rd}$ contacts, respectively.

**Acknowledgements**   We acknowledge the hospitality and assistance in Akademgorodok of Professor Alphya Nesterenko of the State University of Nobosibirsk and Dr. Igor Nesterenko of the Budker Institute of Nuclear Physics of the Russian Academy of Sciences. We also acknowledge the collaboration and assistance of Dr. Bryce A. Babcock throughout our observations at Kastellorizo and Akademgorodok. We are grateful for the assistance of John Antoniadis in the analysis used in Figure 6 and Professor Evaggelos Vanidhis of the Aristotle University in Thessaloniki, Greece, for help with the alignment of the optical instrumentation. We would also like to thank Marek Demianski, William G. Wagner, Eric Kutner, Jeffrey Kutner, Spyros Kanouras, George Pistikoudis and Kostas Tziotziou for logistical assistance. We benefited greatly from the comments of a referee. JMP and the Williams College Eclipse Expedition were funded by grants from the Heliospheric Division of the U.S. National Science Foundation and their Small Grants for Exploratory Research program, NASA, and the Committee for Research and Exploration of the National Geographic Society.